\documentclass[%
 reprint,
 superscriptaddress,
showpacs,preprintnumbers,
 amsmath,amssymb,
 aps,
 prb,
 longbibliography,
]{revtex4-1}

\usepackage{graphicx}
\usepackage{dcolumn}
\usepackage{bm}
\usepackage{color}
\usepackage{ulem}

\begin{document}

\preprint{APS/123-QED}

\title{
Exotic magnetic phases in an Ising-spin Kondo lattice model on a kagome lattice
}

\author{Hiroaki Ishizuka~\cite{KITP}}
\affiliation{
Department of Applied Physics, University of Tokyo, Hongo, 7-3-1, Bunkyo, Tokyo 113-8656, Japan
}

\author{Yukitoshi Motome}
\affiliation{
Department of Applied Physics, University of Tokyo, Hongo, 7-3-1, Bunkyo, Tokyo 113-8656, Japan
}

\date{\today}

\begin{abstract}
Magnetic and electronic states of an Ising-spin Kondo lattice model on a kagome lattice are investigated by a Monte Carlo simulation. In addition to the conventional ferromagnetic and ferrimagnetic orders, we show that this model exhibits several thermally-induced phases, such as partially disordered, Kosterlitz-Thouless-like, and loop-liquid states. In the partially disordered state, we show that the magnetic transition is associated with the charge-gap formation. We find that the density of state shows characteristic peaks reflecting the underlying spin texture. On the other hand, in the loop-liquid state, the formation of closed loops of the same spin sites manifests itself in the peaks in the density of states and the optical conductivity. Our results elucidate the peculiar cooperation between thermal fluctuations and the spin-charge interplay in this frustrated itinerant electron system.
\end{abstract}

\maketitle

\section{Introduction}

The studies on exotic magnetism in geometrically frustrated systems is a hot topic in condensed matter physics, powered by the discovery of many new candidate materials and the development of new theoretical techniques.~\cite{Diep2004,Lacroix2011} A key feature of frustrated magnets is the suppression of conventional magnetic ordering. In these systems, competition between magnetic interactions due to underlying lattice geometry suppresses the formation of magnetic long-range order (LRO), often leaving the system disordered down to zero temperature. In classical spin systems, the disordered ground state is associated with macroscopic degeneracy comprised of spin states that satisfy a local constraint.~\cite{Pauling1935,Wannier1950,Houtappel1950} Such degenerated ground states are extremely sensitive to perturbations, such as subdominant interactions and fluctuations, providing a fertile ground for exotic magnetism.  

\begin{figure}
   \includegraphics[width=\linewidth]{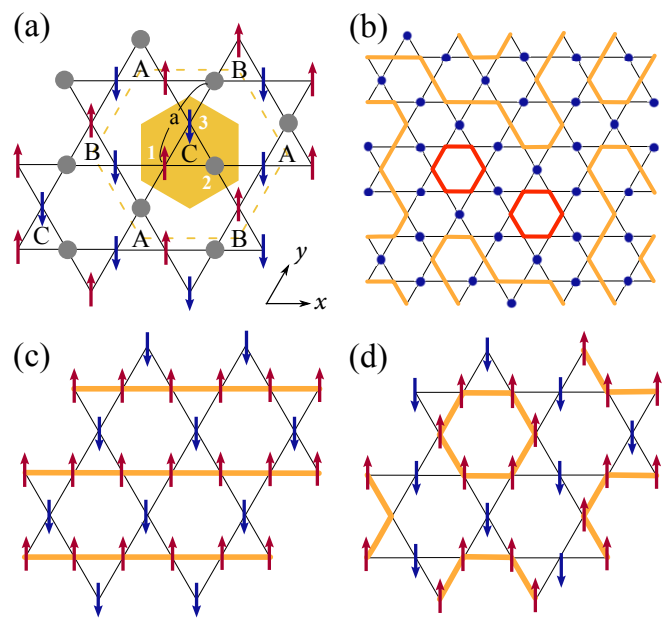}
   \caption{(color online).
   Schematic pictures of (a) partial disorder (PD), (b) loop liquid (LL), (c) $q=0$ ferrimagnetic (FR) order, and (d) $\sqrt3\times\sqrt3$ FR order. The arrows in (a), (c), and (d) indicate ordered Ising moments, and the filled circles in (a) indicate paramagnetic sites. In (a), the shaded hexagon shows the crystallographic unit cell of the kagome lattice (1, 2, and 3 denote the three sublattices), and the dotted hexagon is the magnetic unit cell for the PD state (A, B, and C indicates magnetic sublattices); $a=1$ is the lattice constant. In (b)-(d), the thick lines connect sites with up spins; in (b), the dots denote the sites with down spins and the 6 site loops are colored in red.
   }
   \label{fig:kagome}
\end{figure}

A fundamental, interesting example is found in Ising antiferromagnets on geometrically frustrated lattices. For example, previous studies on the triangular~\cite{Takayama1983,Fujiki1984,Fujiki1986} and kagome lattice~\cite{Takagi1993,Takagi1995} Ising antiferromagnets have reported that Kosterlitz-Thouless (KT) phases are induced by further-neighbor interactions. Another prominent phenomenon caused by subdominant interactions is the partial lifting of the ground-state manifold, such as partially disordered (PD) states.~\cite{Mekata1977,Todoroki2004} The PD states are peculiar magnetic orders characterized by coexistence of magnetically-ordered and paramagnetic (PM) sites forming a periodic structure; an example on the kagome lattice is shown in Fig.~\ref{fig:kagome}(a). Due to the presence of the PM moments, the PD states retain the residual entropy. Hence, they are interpreted as a partial lifting of the degenerate ground-state manifold.

Another interesting example is in itinerant magnets on frustrated lattices.~\cite{Yoshida2011,Matsuda2012,Takatsu2010,Takatsu2014} Some of them are described by the Kondo lattice model, in which itinerant electrons are coupled with localized moments via the local exchange interaction. In this system, the effective interactions between the localized moments are induced by the kinetic motion of itinerant electrons, known as the Ruderman-Kittel-Kasuya-Yosida (RKKY) interaction.~\cite{Ruderman1954,Kasuya1956,Yosida1957} Such effective interactions are, in general, long-ranged with oscillating sign, potentially leading to competing interactions. Meanwhile, in the itinerant electron systems, magnetic LRO may be stabilized by the formation of energy gap in the electronic band structure. In the itinerant magnets, these effects may give rise to frustration and exotic magnetic states. Indeed, recent numerical studies on a triangular lattice model have reported a rich phase diagram. In the case of the model with Ising localized moments, PD and KT-like states were obtained along with several magnetically LRO states.~\cite{Ishizuka2012,Ishizuka2013} On the other hand, a peculiar noncoplanar LRO with scalar chirality ordering was reported in the case of Heisenberg-type localized moments.~\cite{Martin2008,Akagi2010,Kumar2010,Kato2010} We also note that a recent experiment on Ag$_2$CrO$_2$ reported possibility of a PD state.~\cite{Matsuda2012}

In addition, it is also interesting to see how the exotic magnetism affects itinerant electrons. For instance, local correlations in the onsite potentials are known to induce delocalization of electrons, e.g., by a dimer correlation in one dimension~\cite{Dunlap1990} and by formation of loops~\cite{Ishizuka2011} in three dimensions. Also, charge-gap formation by local correlations was reported in a kagome lattice model.~\cite{Ishizuka2013d,Chern2012} Thus, frustrated itinerant magnets are candidates for exploring novel phenomena induced by the cooperation of spin-charge coupling and geometrical frustration. 

Recently, the authors have reported that the Kondo lattice model with Ising localized moments on a kagome lattice shows peculiar magnetic states, such as PD~\cite{Ishizuka2013b} and loop-liquid (LL)~\cite{Ishizuka2013c} phases. However, detailed investigation on the magnetic phase diagrams and their electronic properties have not been reported yet. In this paper, we show comprehensive numerical results on both the magnetic and electronic properties. By using a Monte Carlo (MC) simulation, we show that the model exhibits rich phase diagram with various thermally-induced magnetic phases: KT-like, partially-ferromagnetic (PFM), PD, and LL states, along with conventional magnetic orders such as ferromagnetic (FM) and $q=0$ and $\sqrt3\times\sqrt3$ ferrimagnetic (FR) orders. We also show peculiar electronic and transport properties of the PD and LL states, reflecting the magnetic textures of the underlying localized spins. These results imply that the transport measurements can be used as experimental probes to detect the exotic magnetic states.

The organization of this paper is as follows. In Sec.~\ref{sec:m&m}, we introduce the model and method, with the definitions of physical quantities calculated. In Sec.~\ref{sec:mcresult}, we present our main results on the thermally-induced phases: the magnetic phase diagram, the detailed results of physical quantities, and the electronic and transport properties. Section~\ref{sec:summary} is devoted to discussions and summary. In appendices~\ref{ssec:fm&pfm} and \ref{ssec:sl}, we present MC results on the conventional magnetic phases to complement Sec.~\ref{sec:mcresult}.

\section{Model and Method}\label{sec:m&m}

In this section, we introduce the model and method we used. In Sec.~\ref{ssec:m&m:model}, the Hamiltonian is given for the Kondo lattice model we study in this paper. We present the Monte Carlo (MC) method briefly in Sec.~\ref{sec:m&m:mc} and the definitions of physical quantities in Sec.~\ref{ssec:m&m:quant}. We describe the variational method for the ground state phase diagram in Sec.~\ref{ssec:m&m:variational}.

\subsection{Model}\label{ssec:m&m:model}

We consider a single-band Kondo lattice model on a kagome lattice with localized Ising spin moments. The Hamiltonian is given by
\begin{eqnarray}
H = -t \! \sum_{\langle i,j \rangle, \sigma} \! ( c^\dagger_{i\sigma} c_{j\sigma} + \text{H.c.} ) - J \sum_{i}\sigma_i^z S_i.
\label{eq:H}
\end{eqnarray}
Here, the first term represents hopping of itinerant electrons, where $c_{i\sigma}$ ($c^\dagger_{i\sigma}$) is the annihilation (creation) operator of an itinerant electron with spin $\sigma= \uparrow, \downarrow$ at $i$th site, and $t$ is the transfer integral. The sum $\langle i,j \rangle$ is taken over NN sites on the kagome lattice. The second term is the onsite interaction between localized spins and itinerant electrons, where $\sigma_i^z = c_{i\uparrow}^\dagger c_{i\uparrow} -  c_{i\downarrow}^\dagger c_{i\downarrow}$ corresponds to the $z$-component of itinerant electron spin, and $S_i = \pm 1$ is the localized Ising moment at $i$th site; $J$ is the coupling constant (the sign of $J$ does not matter in the present model). Hereafter, we take $t=1$ as the unit of energy, and the lattice constant $a = 1$ [see Fig.~\ref{fig:kagome}(a)]. 

\subsection{Monte Carlo method}\label{sec:m&m:mc}

To investigate thermodynamic properties of the model~(\ref{eq:H}), we utilized a MC simulation which has been widely applied to similar models.~\cite{Yunoki1998} Although there have already been numbers of papers which describe this method, here we briefly review the MC technique to make the paper self-contained.

The model we study in this paper, Eq.~(\ref{eq:H}), belongs to the class of models in which fermions are coupled to classical fields. The partition function is given by
\begin{eqnarray}
Z={\rm Tr}_f{\rm Tr}_c \exp[-\beta(H-\mu\hat{N_e})],
\label{eq:Z}
\end{eqnarray}
where ${\rm Tr}_f$ is the trace over classical degree of freedom (in the current case, Ising spin configurations), and ${\rm Tr}_c$ is the trace over itinerant fermions. Here, $\beta=1/T$ is the inverse temperature (we set the Boltzmann constant $k_{\rm B}=1$), $\mu$ is the chemical potential, and $\hat{N_e}$ is the total number operator for fermions.

The key feature of the Hamiltonian in Eq.~(\ref{eq:H}) is that the Hamiltonian is block diagonal for different spin configurations $\{S_i\}$, i.e., the two traces in Eq.~(\ref{eq:Z}) are taken separately. Hence, one can calculate the partition function by estimating the trace ${\rm Tr}_f$ by a classical MC sampling over $\{S_i\}$ using the Markov-chain MC method. The MC weight for a given $\{S_i\}$ is calculated by taking the fermion trace ${\rm Tr}_c$ in the following form, 
\begin{eqnarray}
P_{\rm MC}(\{S_i \}) = \exp[ -S_{\rm eff}(\{S_i \}) ],
\label{eq:P}
\end{eqnarray}
where $S_{\rm eff}$ is the effective action calculated as
\begin{align}
S_{\rm eff}(\{S_i \})= - \sum_\nu \log[1 + \exp\{-\beta(E_\nu(\{S_i \})-\mu)\}].
\label{eq:S_eff}
\end{align}
Here, $E_\nu(\{S_i \})$ are the energy eigenvalues for the configuration $\{S_i \}$, which are readily calculated by the exact diagonalization as it is a one-body problem in a static potential.

For the MC sampling, we utilized single-spin flip update with the standard METROPOLIS algorithm. On the other hand, to overcome the freezing of MC sampling in the LL state, we also implemented the loop-update method~\cite{Rahman1972,Barkema1998,Melko2001} in our calculations. Also, some of the low-temperature data were calculated starting from a mixed initial spin configuration of low-temperature ordered and high-temperature disordered states.~\cite{Ozeki2003}

The calculations were conducted up to the system size $N=3\times N_{\rm s}$ with $N_{\rm s}=9 \times 9$ under the periodic boundary conditions. Thermal averages of physical quantities were calculated for typically 15000-80000 MC steps after 5000-18000 steps for thermalization. The results are shown in the temperature range where the acceptance ratio is larger than  $\sim$1\%. We divided the MC measurements into five bins and estimated the statistical errors by the standard deviations among the bins.

\subsection{Physical quantities}\label{ssec:m&m:quant}

In this section, we introduce the definitions of physical quantities we calculated in the MC simulation. In this study, we calculate both electronic and magnetic properties of the model in Eq.~(\ref{eq:H}). The electronic properties, such as the density of states (DOS) and the optical conductivity, are computed for the itinerant electrons. On the other hand, for the magnetic properties, we focus on the contribution from the localized moments, which is simple to compute in the MC simulation, and ignore that from the itinerant electrons. This is justified when the magnitude of the localized spins is much larger than $1/2$, such as in some of rare-earth magnets. It is also justified when the coupling $J$ is strong and the itinerant electron spins are almost fully polarized along the localized moments.

In the Monte Carlo simulation, the formation of a magnetic long-range order (LRO) is detected by the spin structure factor for the Ising spins,
\begin{eqnarray}
S({\bf q}) = \frac{1}{N} \sum_\alpha \sum_{i,j\in\alpha} \langle S_i S_j \rangle \exp({\rm i} {\bf q}\cdot{\bf r}_{ij}),
\label{eq:Sq}
\end{eqnarray}
where the bracket denotes the thermal average in the grand canonical ensemble, and ${\bf r}_{ij}$ is the position vector from $i$ to $j$th site. $\alpha=1,2,3$ denotes the three sublattices [see Fig.~\ref{fig:kagome}(a)], and the sum of $i$ and $j$ is taken over all the sites belonging to the same sublattice. 

In the following, we show that the model in Eq.~(\ref{eq:H}) exhibits a variety of phases with different magnetic orders. Among them, the $\sqrt3\times\sqrt3$ ferrimagnetic (FR) order shown in Fig.~\ref{fig:kagome}(c) is signaled by coexisting peaks of $S({\bf q})$ at ${\bf q}={\bf 0}$ and ${\bf q}=\pm(2\pi/3,-2\pi/3)$. The Bragg peaks at ${\bf q}=\pm(2\pi/3,-2\pi/3)$ also appear for the PD state in Fig.~\ref{fig:kagome}(a),  but the peak at ${\bf q}={\bf 0}$ is absent in this phase. On the other hand, the simple FM and $q=0$ FR [Fig.~\ref{fig:kagome}(d)] orders develop a peak only at ${\bf q}={\bf 0}$. These two phases are distinguished by the net magnetization
\begin{eqnarray}
m = \langle M^2\rangle^{1/2},
\end{eqnarray}
where
\begin{eqnarray}
M = \frac1N\sum_i S_i.
\end{eqnarray}
Here, $m$ takes $1$ and $1/3$ for FM and $q=0$ FR states, respectively. We also calculate the susceptibility of $M$ by the fluctuation formula
\begin{eqnarray}
\chi = \frac{N}{T} \{m^2 - \langle\left|M\right|\rangle^2 \}.
\end{eqnarray}

In the KT-like state appearing in our model, in principle, no Bragg peaks develop as it is a quasi-LRO. However, in finite-size calculations, it is difficult to distinguish the quasi-LRO from true LRO solely by the structure factor, as the correlation length in the KT state is divergent and exceeds the system size. To discriminate the KT-like state, it is helpful to use the pseudospin defined for each three-site unit cell,~\cite{Takayama1983,Fujiki1984}
\begin{eqnarray}
\tilde{\bf S}_l = 
\left(
\begin{array}{ccc}
\frac2{\sqrt6} & -\frac1{\sqrt6} & -\frac1{\sqrt6} \\
0              &  \frac1{\sqrt2} & -\frac1{\sqrt2} \\
\frac1{\sqrt3} &  \frac1{\sqrt3} &  \frac1{\sqrt3} \\
\end{array}
\right)
\left(
\begin{array}{c}
S_i  \\
S_j  \\
S_k  \\
\end{array}
\right),
\end{eqnarray}
where $l$ is the index for the three-site crystalographic unit cells, and ($i, j, k$) denotes the three sites in the $l$th unit cell. Taking the summation for each magnetic sublattice, we define
\begin{eqnarray}
\tilde{\bf M}^a = \frac{3}{N_{\rm s}} \sum_{l\in a} \tilde{\bf S}_l
\end{eqnarray}
where $a=$A,B,C denotes the magnetic sublattices shown in Fig.~\ref{fig:kagome}(a). The quantity $\tilde{\bf M}^a$ is useful in identifying local spin correlations.

In the MC simulation, we calculate 
\begin{eqnarray}
M^a_{xy} &=& \langle \{(\tilde{M}^a_x)^2 + (\tilde{M}^a_y)^2\}^{1/2} \rangle, 
\label{eq:Mxy} \\
M^a_z &=& \langle |\tilde{M}^a_z| \rangle,
\label{eq:Mz}
\end{eqnarray}
and the corresponding susceptibilities,
\begin{eqnarray}
\chi^a_{xy} &=& \frac{N}{T} \{\langle (\tilde{M}^a_x)^2 + (\tilde{M}^a_y)^2 \rangle - (M^a_{xy})^2 \}, \\
\chi^a_z &=& \frac{N}{T} \{\langle (\tilde{M}^a_z)^2 \rangle - (M^a_z)^2 \},
\end{eqnarray}
where $\tilde{\bf M}^a = (\tilde{M}^a_x, \tilde{M}^a_y, \tilde{M}^a_z)$.

To distinguish the KT-like state from other LROs, especially from FR and PD, we use the azimuth parameter of $\tilde{{\bf M}}^{a}$ defined by~\cite{Todoroki2004,Ishizuka2012}
\begin{eqnarray}
\psi^a = ({\cal M}^a)^3 
\cos{6 \phi_{M^a}},
\label{eq:psi}
\end{eqnarray}
where $\phi_{M^a}$ is the azimuth of $\tilde{\bf M}^a$ in the $xy$ plane and ${\cal M}^a = \frac38 (M^a_{xy})^2$.
The parameter $\psi^a$ has a negative value and approaches $\psi^a \to -\frac{27}{64}$ as $T\to 0$ for the perfect PD ordering, while it becomes positive and approaches $\psi^a \to 1$ for the perfect FR ordering; $\psi^a=0$ for both PM and KT-like phases in the thermodynamic limit $N \to \infty$. In previous studies for the models on a triangluar lattice,~\cite{Ishizuka2012,Ishizuka2013} it was shown that a similar parameter to $\psi^a$ shows much faster convergence to the value in the thermodynamic limit with increasing system sizes compared to other quantities, such as the structure factor. Hence, it is useful for distinguishing the PD from the KT-like state.

In all the states we discuss in the following, the quantities $M_{xy}^a$, $M_z^a$, $\chi_{xy}^a$, $\chi_z^a$, and $\psi^a$ are essentially the same for all the sublattices, $a={\rm A}$, B, and C. Hence, we will show the averages over the sublattices in the following results.

In addition, we measure local spin correlations in each triangle by
\begin{eqnarray}
p_{\alpha^\prime} =\left\{
\begin{array}{rl}
 1 \qquad& \text{for two-up one-down}\\
-1 \qquad& \text{for one-up two-down}\\
 0 \qquad& \text{otherwise,}
\end{array}
\right.
\end{eqnarray}
where $\alpha^\prime$ is the index for the three-site triangles (both the upward and downward triangles) in the kagome lattice. In the MC simulation, we calculate the probability $P$ that the triangles are in two-up one-down (or one-up two-down) coherently, which is calculated as 
\begin{eqnarray}
P = \Big\langle\Big(\frac{1}{2N_{\rm s}}\sum_{\alpha^\prime} p_{\alpha^\prime} \Big)^2\Big\rangle^{1/2}.
\end{eqnarray}
As we will show in Sec.~\ref{ssec:ll}, this parameter is useful in distinguishing the LL state, which is characterized by development of local correlation. The susceptibility of $P$ is also calculated using the fluctuation formula,
\begin{eqnarray}
\chi_P = \frac{N_{\rm s}}T \left\{ P^2 - \Big\langle\Big|\frac{1}{2N_{\rm s}}\sum_{\alpha^\prime} p_{\alpha^\prime} \Big|\Big\rangle^2 \right\}.
\end{eqnarray}

In the MC simulation, we calculate the temperature dependence of these quantities at a fixed electron density in the grand canonical ensemble. The temperature dependence of chemical potential, $\mu(T)$, for a given electron density is determined by calculating the electron density
\begin{eqnarray}
n=\frac1N \sum_{i\sigma} \langle c_{i\sigma}^\dagger c_{i\sigma} \rangle,
\end{eqnarray}
while tuning $\mu$ as a function of temperature. In the following MC results, the error for $n$ is typically $\sim 0.001$.

We also calculate the optical conductivity of itinerant electrons in this model by the standard Kubo formula. In the present system, the expectation value of the electric current operator in the Kubo formula is also diagonal in terms of the Ising spin configurations, and hence, calculated as
\begin{eqnarray}
&&
\langle \psi_f^{\nu},\{S_i\} | J_{\hat{\bf n}} |\psi_f^{\nu^\prime},\{S_i\}^\prime \rangle =\nonumber\\
&&\qquad\delta_{\{S_i\},\{S_i\}^\prime} \langle \psi_f^{\nu},\{S_i\} | J_{\hat{\bf n}} |\psi_f^{\nu^\prime},\{S_i\} \rangle, \label{eq:bracket}
\end{eqnarray}
where $J_{\hat{\bf n}}$ is the current operator in the $\hat{\bf n}$ direction, $\hat{\bf n}$ is a normalized vector in $xy$ plane, and $\delta_{\{S_i\},\{S_i\}^\prime} = 1$ if $\{S_i\}=\{S_i\}^\prime$, while otherwise $0$. Here, $|\psi_f^{(\nu)},\{S_i\}\rangle$ is the $\nu$th single-particle eigenstate with spin configuration $\{S_i\}$. Using Eq.~(\ref{eq:bracket}), the optical conductivity is calculated as
\begin{eqnarray}
\sigma(\omega) = \langle\sigma^{\{S_i\}}(\omega)\rangle = \sum_{\{S_i\}} P(\{S_i\})\sigma^{\{S_i\}}(\omega)
\end{eqnarray}
where $\sigma^{\{S_i\}}(\omega)$ is the optical conductivity for spin configuration $\{S_i\}$, given in the form
\begin{align}
\sigma^{\{S_i\}}
(\omega) = -\mathrm{i} \sum_{\alpha,\alpha^\prime} \frac{f(\varepsilon_\alpha) - f(\varepsilon_{\alpha^\prime})}{\varepsilon_\alpha-\varepsilon_{\alpha^\prime}} \frac{\left< \alpha\right| J_{\hat{\bf n}} \left|\alpha^\prime\right> \left< \alpha^\prime\right| J_{\hat{\bf n}} \left|\alpha\right>}{\omega + \varepsilon_\alpha - \varepsilon_{\alpha^\prime} + \mathrm{i}\tau}.\;\label{eq:kubo_part}
\end{align}
Here, the indices of energy and eigenstate are abbreviated for simplicity: $\varepsilon_\alpha = \varepsilon_\alpha(\{S_i\})$ is the eigenenergy for $\alpha$th one particle state $\left|\alpha\right> = \left|\alpha,\{S_i\}\right>$. In the results shown below, we calculate the mean average for the conductivity measured along and perpendecular to $x$ directions. The axes are defined as shown in Fig.~\ref{fig:kagome}(a). Also, we set $e^2/h=1$, where $e$ is elementary charge, $h$ is Planck's constant and take $\tau=0.01$. Hence, the optical conductivity is calculated by taking MC average of Eq.~(\ref{eq:kubo_part}) over different spin configurations.

\subsection{Variational calculation}\label{ssec:m&m:variational}

In addition to the MC simulation, we also investigate the ground state using a variational calculation, which was used in the previous study of the triangular lattice model.~\cite{Ishizuka2013c} The ground state phase diagram is obtained by comparing the ground state energy for different LRO spin configurations found in the MC simulation.
The phase separated regions at $T=0$ are also identified by the variational method from the jumps of $n$ at the magnetic phase transitions while changing the chemical potential $\mu$. See Ref.~\onlinecite{Ishizuka2013c} for details.

\section{Monte Carlo Results}\label{sec:mcresult}

In this section, we present our MC results on the model in Eq.~(\ref{eq:H}). In Sec.~\ref{ssec:diagram}, we discuss the phase diagram of the model obtained by the MC simulation focusing on the thermally-induced phases: the partially disordered and the loop-liquid phases. Details of these phases and their transport properties are discussed in Secs.~\ref{ssec:pd} and \ref{ssec:ll}. For the MC results of ferromagnetic and ferrimagnetic phases, see Appendices~\ref{ssec:fm&pfm} and \ref{ssec:sl}.

\subsection{Phase diagram}\label{ssec:diagram}

\begin{figure}
   \includegraphics[width=.92\linewidth]{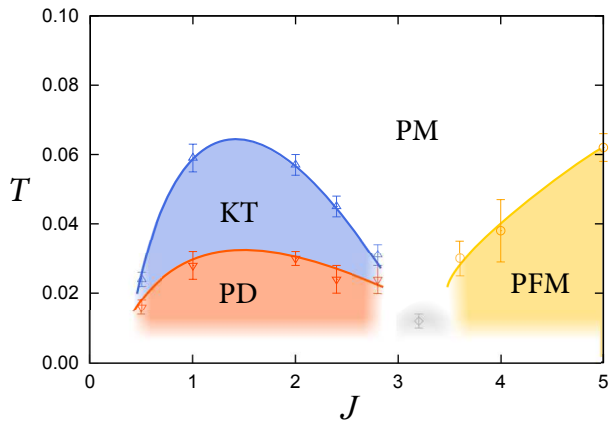}
   \caption{(color online).
   Phase diagram of the model in Eq.~(\ref{eq:H}) at $n=2/3$ obtained by the MC simulation. The symbols shows the critical temperatures $T_c$ for magnetic states: partial disorder (PD), partially ferromagnetic (PFM), Kosterlitz-Thouless-like (KT), and paramagnetic (PM) states. The critical temperature for the KT-like state, $T_{\rm KT}$, is estimated by the extrapolation of the peak of $\chi_{xy}^a$, while that for PD by extrapolation of $\psi^a$. $T_c$ for the PFM state shown is estimated from the Binder analysis of $m$. The gray zone at $J\sim 3.2$ shows the phase separation. The curves connecting the symbols are guides for the eye. See Secs.~\ref{ssec:pd} and \ref{ssec:fm&pfm} for details.
   }
   \label{fig:diagram_n23}
\end{figure}

We first start from the phase diagram at $n=2/3$ calculated by the MC simulation. Figure~\ref{fig:diagram_n23} shows the phase diagram at $n=2/3$ with varying $J$. In the low-temperature region, the phase diagram is dominated by two different magnetic phases: PD and PFM  states. In the PD state, the spins form a three-sublattice magnetic superstructure as shown in Fig.~\ref{fig:kagome}(a). Here, two out of three spins in the crystallographic unit cell are antiferromagnetically ordered while the remaining site is still PM. In the neighboring unit cells, this magnetic unit is rotated by $\frac{2\pi}3$, forming a magnetic unit cell with nine sites [see Fig.~\ref{fig:kagome}(a)]. On the other hand, in the PFM state, the system exhibits a FM order whose ordered moment saturates at a smaller value than the full polarization as temperature is lowered.

In our results in Fig.~\ref{fig:diagram_n23}, the two phases are separated by a small region of phase separation, that appears slightly above $J=3$; the PD state is found for $J\lesssim 3.0$ and PFM state for $J\gtrsim 3.4$. In the range of $3.0 \lesssim J\lesssim 3.4$, the MC calculation for $n$ becomes unstable as $T\to 0$: it is hard to tune the chemical potential $\mu$ to fix $n$ at $2/3$. This is a signal of phase separation. The gray point in the phase diagram at $J=3.2$ shows the temperature where the calculation of $n$ becomes unstable for $N_{\rm s}=9\times9$.

While increasing temperature, the PFM state shows a second-order phase transition to the PM state. On the other hand, for the PD state, another phase appears in between the PD and high-temperature PM phases. This phase does not show a clear indication of LRO, but shows characteristic behavior that resembles the KT phase. Hence, we call this phase the KT-like phase.

\begin{figure}
   \includegraphics[width=\linewidth]{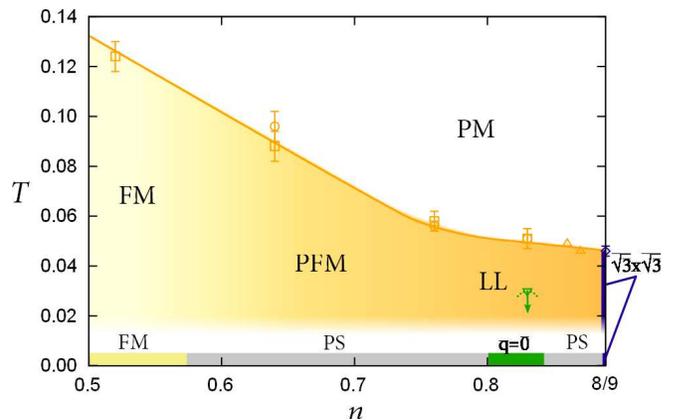}
   \caption{(color online).
	   Phase diagram of the model in Eq.~(\ref{eq:H}) at $J=6$ obtained by the MC simulation. The symbols shows the critical temperatures $T_c$ for magnetic states: ferromagnetic (FM), partially ferromagnetic (PFM), loop liquid (LL), $q=0$ ferrimagnetic ($q=0$), and $\sqrt3\times \sqrt3$ ferrimagnetic ($\sqrt3\times \sqrt3$) states. $T_c$ for the $\sqrt3\times \sqrt3$ state at $n=8/9$ is shown by the diamond, which is determined by the extrapolation of the peak of $\chi_{xy}^a$.  Meanwhile, the upper limit for $T_c$ for the $q=0$ state at $n=0.83$ is shown by the downward triangle, which is given by the temperature we reached with $N_{\rm s}=8^2$ calculations. The squares (circles) show $T_c$ determined from the Binder analysis of $m$ ($P$), and the upward triangles show $T_c$ determined by the extrapolation of the peak of $\chi_{m}$. The curve connecting the symbols is a guide for the eye. The strip at the bottom is the ground state phase diagram obtained by the variational calculation for three magnetic orders, FM, $q=0$, and $\sqrt3\times \sqrt3$. PS is the phase separation between the neighboring two phases. See Secs.~\ref{ssec:pd}, \ref{ssec:ll}, and Appendices~\ref{ssec:fm&pfm} and \ref{ssec:sl} for details.
   }
   \label{fig:ntdiag}
\end{figure}

We next show the magnetic phase diagram while changing the electron density $n$. Figure~\ref{fig:ntdiag} shows the MC result at $J=6$. When $J=6$, at all the values of $n<8/9$ we calculated, the system exhibits a phase transition characterized by the development of a net magnetization $m$ with no magnetic superlattice structure. When the electron density is sufficiently small, the ground state is the FM ordered state and $m$ approaches $1$ as $T\to 0$. However, with increasing $n$, the saturation value of $m$ gradually decreases from 1, which is shown as the PFM in Fig.~\ref{fig:ntdiag}. With further increasing $n$, it approaches $m=1/3$ around $n\sim0.8$. For $0.8 \lesssim n < 8/9$, the system is in the loop-liquid (LL) state, in which all the triangles in the kagome lattice takes two-up one-down configurations, as shown in Fig.~\ref{fig:kagome}(b). In this phase, the loops of up-spins separated by down spins are thermally fluctuating between different loop patterns.

In the vicinity of this LL state, as decreasing temperature or as further increasing $n$, the LL state exhibits phase transitions to ferrimagnetic LROs. In our MC simulation, we identify two different transitions; one is the transition to the state with $q=0$ LRO of two-up one-down spin configurations [Fig.~\ref{fig:kagome}(d)], and the other to the state with $\sqrt3\times\sqrt3$ LRO [Fig.~\ref{fig:kagome}(c)]. The former is observed while decreasing temperature at $n\sim 0.83$, and the latter is found by increasing $n$ to a commensurate filling $n=8/9$. In the corresponding density regions, the two phases are obtained in the variational calculation for the ground state, as shown in the strip at the bottom of Fig.~\ref{fig:ntdiag}.

These two LRO states can be viewed as the ``solidification" of the emergent loops in the two extreme cases; the former is a periodic array of one-dimensional chains, while the latter the shortest six-site hexagons. Interestingly, the LL state extends in the density region between these two ``crystal" phases. In the phase diagram, in principle, the phase boundary between the LL state and the two FR states are well defined by the formation of magnetic LROs. On the other hand, the transition between FM, PFM, and LL states are crossover.

\subsection{Partially disordered phase}\label{ssec:pd}

In this section, we present the MC results for the PD state in the phase diagram in Fig.~\ref{fig:diagram_n23}. We discuss the magnetic properties in Sec.~\ref{sssec:pd_mc} and the electronic structure in Sec.~\ref{sssec:pd_opt}.

\subsubsection{Magnetic properties} \label{sssec:pd_mc}

\begin{figure}
   \includegraphics[width=0.8\linewidth]{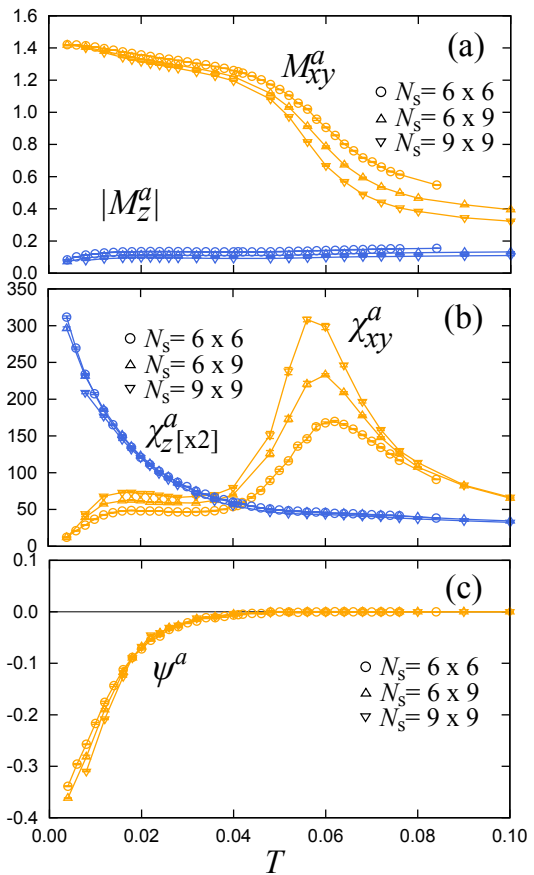}
   \caption{(color online).
   MC results for the PD state at $n=2/3$ and $J=2$: (a) $M_{xy}^a$ and $|M_z^a|$, (b) $\chi_{xy}^a$ and $\chi_{z}^a$, and (c) $\psi^a$. The data are calculated for the system sizes $N_{\rm s}= 6 \times 6$, $6\times 9$, and $ 9\times 9$.
   }
   \label{fig:mc_n23pd}
\end{figure}

Figure~\ref{fig:mc_n23pd} shows the results of the MC calculation at $n=2/3$ and $J=2$. As shown in Fig.~\ref{fig:mc_n23pd}(a), $M_{xy}^a$ monotonically increases as temperature is decreased, with showing a pronounced increase at $T \sim 0.06$. In addition, it exhibits a small shoulder at $T \sim 0.015$ before approaching the value at the lowest temperature. The two anomalies are more clearly observed in the corresponding susceptibility $\chi_{xy}^a$ plotted in Fig.~\ref{fig:mc_n23pd}(b); $\chi_{xy}^a$ shows a peak which grows as the system size increases at $T \sim 0.06$ and a hump structure at $T \sim 0.015$. These results imply the presence of two successive transitions. Indeed, these are phase transitions as discussed below; their critical temperatures are estimated as $T_{\rm KT}=0.057(3)$ and $T_c^{\rm PD}=0.030(2)$ by the finite-size analyses of $\chi_{xy}^a$ and the azimuth parameter $\psi^a$, respectively.

\begin{figure}
   \includegraphics[width=\linewidth]{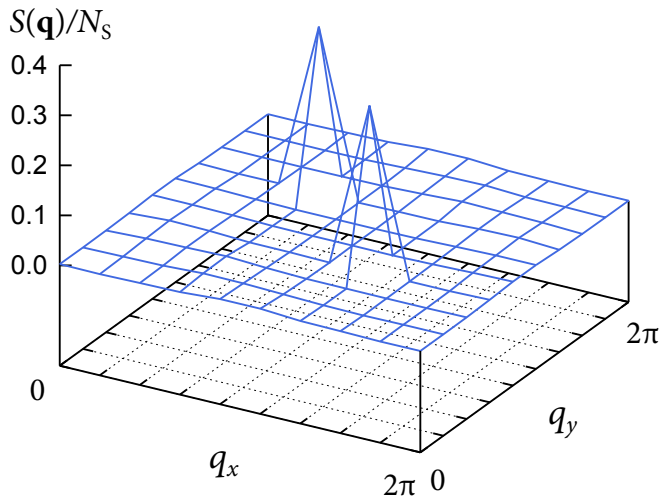}
   \caption{(color online).
   MC results of the spin structure factor $S({\bf q})$ divided by $N_{\rm s}$ at $J=2$, $n=2/3$, and $T=0.004$. The data are calculated for the system size $N_{\rm s}= 9^2$.
   }
   \label{fig:mc_n23_sq}
\end{figure}

\begin{figure}
   \includegraphics[width=\linewidth]{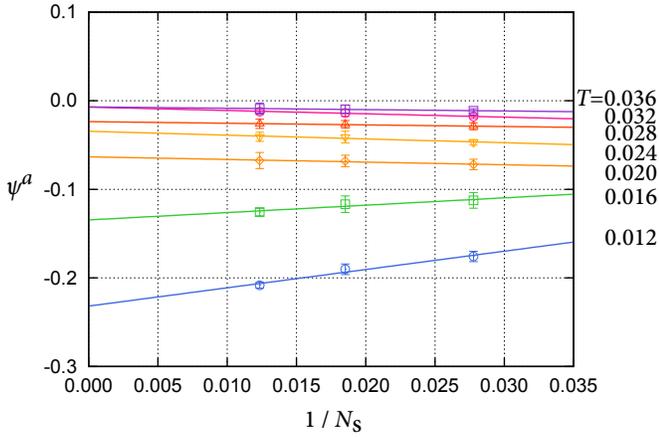}
   \caption{(color online).
   System size dependence of $\psi^a$ at $J=2$ and $n=2/3$. The solid lines indicate linear fittings of the data at each temperature.
   }
   \label{fig:mc_n23_psi}
\end{figure}

Let us first discuss the phase transition at a lower temperature $T_c^{\rm PD}$. In the low-temperature region, $M_{xy}^a$ approaches $\sqrt2$ while $|M_{z}^a|$ is essentially zero in the thermodynamic limit, as shown in Fig.~\ref{fig:mc_n23pd}(a). In addition, the azimuth parameter $\psi^a$ shows a sharp decrease below $T \lesssim 0.03$, from $\psi^a=0$ to $ \sim - 0.4$, as shown in Fig.~\ref{fig:mc_n23pd}(c). The nonzero $\psi^a$ indicates a spontaneous breaking of sixfold rotational symmetry of ${\bf M}^\alpha$, and the negative value approaching $-27/64$ suggests that the system exhibits an instability toward PD.~\cite{Ishizuka2012,Ishizuka2013} In addition, we find that the spin structure factor exhibits the peaks corresponding to the $\sqrt3 \times \sqrt3$ order as shown Fig.~\ref{fig:mc_n23_sq}. From these results, we conclude that the system exhibits the PD state with period $\sqrt3 \times \sqrt3$ below $T \lesssim 0.03$ [see Fig.~\ref{fig:kagome}(a)]. The critical temperature $T_c^{\rm PD}$ for PD ordering is determined by the size extrapolation of $\psi^a$ shown in Fig.~\ref{fig:mc_n23_psi}. We obtain the estimate $T_c^{\rm PD} = 0.030(2)$ by the temperature at which the size-extrapolated value of $\psi^a$ deviates from zero beyond the error bars.

We note that $\chi_z^a$ shows a monotonic increase with decreasing temperature, as shown in Fig.~\ref{fig:mc_n23pd}(b). This is distinct behavior from the conventional antiferromagnetic ordering, which shows a monotonic decrease below $T_c$. This behavior is ascribed to the presence of PM spins, which is consistent with the PD state.

Next, we discuss the phase transition at a higher temperature $T_{\rm KT}$. The transition is signaled by the divergent peak of $\chi_{xy}^a$ and corresponding rapid rise of $M_{xy}^a$. In the intermediate-temperature region $T_c^{\rm PD} < T < T_{\rm KT}$, however, $M_{xy}^a$ exhibits a considerable finite-size effect. On the other hand, $|M_z^a|$ and $\psi^a$ shows almost no change. Furthermore, $\psi^a$ is extrapolated to zero within statistical errors in the limit of $N \to \infty$ (see Fig.~\ref{fig:mc_n23_psi}). This is in contrast to the PD state and the $\sqrt3\times\sqrt3$ long-range ordered state, where $\psi^a$ should become negative and positive, respectively. On the other hand, similar behavior was observed in the KT phase with quasi-LRO in the Ising antiferromagnets on triangular and kagome lattices.~\cite{Takayama1983,Fujiki1984,Fujiki1986,Takagi1993,Takagi1995} Hence, we conclude that the intermediate phase for $T_c^{\rm PD} < T < T_{\rm KT}$ is a KT-like phase. Thus, in the region $J\lesssim3$, the system shows successive phase transitions from the high-temperature PM phase to the KT-like phase, and from the KT-like phase to the PD state.

One point to be noted is that another phase transition is anticipated from the PD state to the true ground state. Although we could not detect such a phase transition within the current MC simulation, it is unlikely that the PD state remains as the ground state since it retains finite residual entropy due to the paramagnetic moments. It is more likely that, in the lowest temperature, the degeneracy will be lifted by the long-range RKKY interactions induced by the spin-charge coupling,~\cite{Ruderman1954,Kasuya1956,Yosida1957} driving a transition to an LRO state. If this is the case, the transition temperature is much lower than we could approach in the current MC simulation, indicating the energy scale of the relevant RKKY interaction is extremely small. We note that a similar feature was also reported recently in an Ising spin Kondo lattice model on a triangular lattice, where the PD state survived down to an extremely low temperature at $n=1/3$.~\cite{Ishizuka2012,Ishizuka2013} On the other hand, another possibility is that the PD state is taken over by a phase separation at the lowest temperature.

%

\subsubsection{Electronic structure}\label{sssec:pd_opt}

\begin{figure}
   \includegraphics[width=\linewidth]{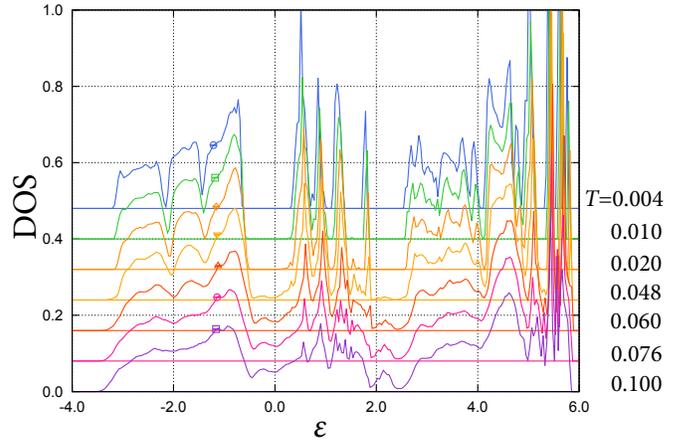}
   \caption{(color online).
   MC results for the DOS calculated while varying temperature at $J=2$ and $n=2/3$. The typical error bars are shown at $\varepsilon\sim -1.2$.
   }
   \label{fig:mc_n23_dos}
\end{figure}

In the previous studies on the PD state in the triangular lattice case,~\cite{Ishizuka2012,Ishizuka2013} it was shown that the energy gap opens in the DOS with the formation of PD. Namely, a metal-insulator transition takes place at the phase transition. It was also argued that the energy gap formation may play a role in stabilizing the PD state. To see whether such behavior also takes place in the current kagome case, we calculated the electronic DOS using the MC calculation. Figure~\ref{fig:mc_n23_dos} shows the MC results of the temperature dependence of the DOS at $J=2$. The Fermi level at $n=2/3$ is set to be $\varepsilon=0$, and typical statistical errors are shown at $\varepsilon\sim -1.2$.

At $T=0.1 > T_c^{\rm PD}$, the DOS shows featureless structure with a nonzero DOS at the Fermi level. In contrast, the data at $T=0.004 < T_c^{\rm PD}$ shows a clear energy gap at the Fermi level, indicating that the PD state is an insulator. This behavior resembles that of the PD state in the triangular lattice model. Hence, it is likely that the energy-gap formation contributes to the stabilization of the PD state also in the current kagome lattice case.  

One difference from the triangular lattice case is that the KT-like state appears above the PD state. In the present calculation, the DOS at the Fermi level appears to decrease below $T \sim T_{\rm KT}=0.057(3)$. As the KT-like state is an intermediate state with quasi-LRO, it is natural that the gap opens at $T=T_c^{\rm PD}$ where the true LRO sets in. Nevertheless, in our calculation, a precursor of the gap formation is observed in the KT-like phase. Because of the finite size effect, however, it is not clear at which temperature a full gap opens in the MC data.

Another notable feature of the DOS is the spikes at $\varepsilon\in[0,2]$. These spikes are likely to originate in the peculiar magnetic texture of the PD state. In the PD state, the six-site antiferromagnetic loops of ordered spins are separated from each other by the paramagnetic sites. Due to the presence of these six-site spin clusters, in a snapshot of the Ising spin configuration, the length of same spin chains are restricted to $L=1,3,5,$ and $6$; $L=3$ and $5$ are open strings of same spin sites, $L=6$ is a closed loop, and $L=1$ is an isolated site. Hence, if $J$ is sufficiently large and the electrons are confined into the strings and loops, they form a set of peaks in the DOS that corresponds to the confined states.

\begin{figure}
   \includegraphics[width=\linewidth]{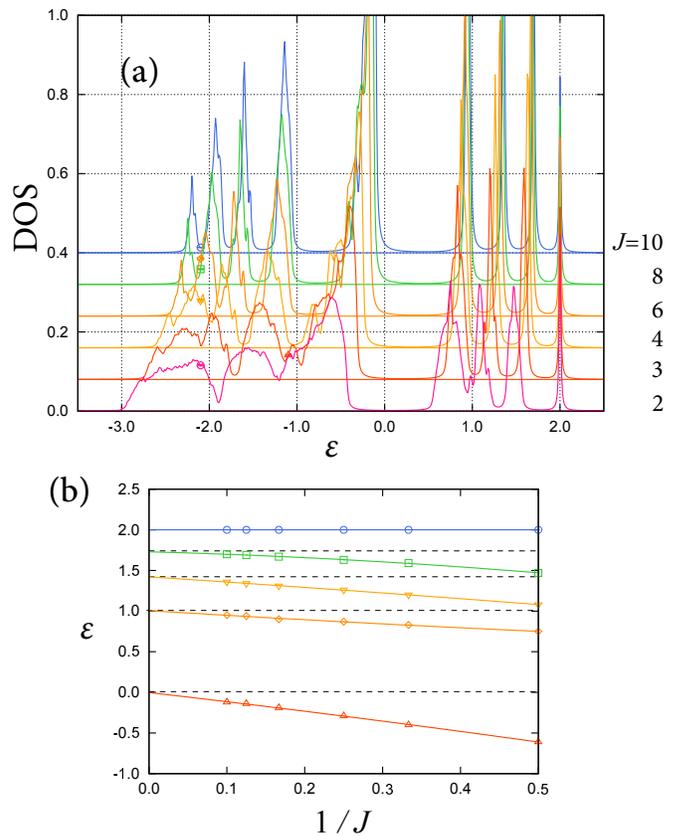}
   \caption{(color online).
   (a) The DOS calculated by simple average over randomly generated PD spin patterns for several values of $J$. The chemical potential is set to $-J$. (b) $J$ dependence of the position of the peaks in (a). See the text for details. The typical error bars are shown at $\varepsilon= -2.11$.
   }
   \label{fig:pd_dos}
\end{figure}

To confirm this scenario, we calculate the DOS for different $J$ by assuming the PD state. The calculations were done by randomly generating the spin configurations in the PD state of $N_{\rm s}=12\times12$. The DOS was calculated by taking $4\times 4$ superlattices of the $N_{\rm s}=12\times12$ unit cell, and averaging over 40 different spin patterns. Figure~\ref{fig:pd_dos}(a) shows the MC result of DOS calculated for different $J$. The results show four peaks in the energy range of $\varepsilon \in [0,2]$ above the $n=2/3$ energy gap. The $J$ dependences of the peak energies are plotted in Fig.~\ref{fig:pd_dos}(b). The solid curves show fittings by $\varepsilon = a + b/J + c/J^2$, and the dotted lines indicate the eigenenergies of the confined states expected for the finite-length strings and loops at $J\to \infty$.~\cite{Jaubert2012} The result shows that the peak energies approach the eigenenergies of the confined states with increasing $J$. Hence, the spikes in Fig.~\ref{fig:mc_n23_dos} are ascribed to nearly confined states in the peculiar spin textures appearing in the PD state on the kagome lattice. This behavior is contrasting to the case of the triangular lattice.~\cite{Ishizuka2012,Ishizuka2013}

\subsection{Loop liquid state} \label{ssec:ll}

In this section, we present the MC results for the LL state in the phase diagram in Fig.~\ref{fig:ntdiag}. We discuss the magnetic properties in Sec.~\ref{sssec:ll_mag}. We also discuss the electronic structure and the optical conductivity in Secs.~\ref{sssec:ll_ele} and \ref{sssec:ll_opt}, respectively.

\subsubsection{Magnetic properties}\label{sssec:ll_mag}

\begin{figure}
   \includegraphics[width=.85\linewidth]{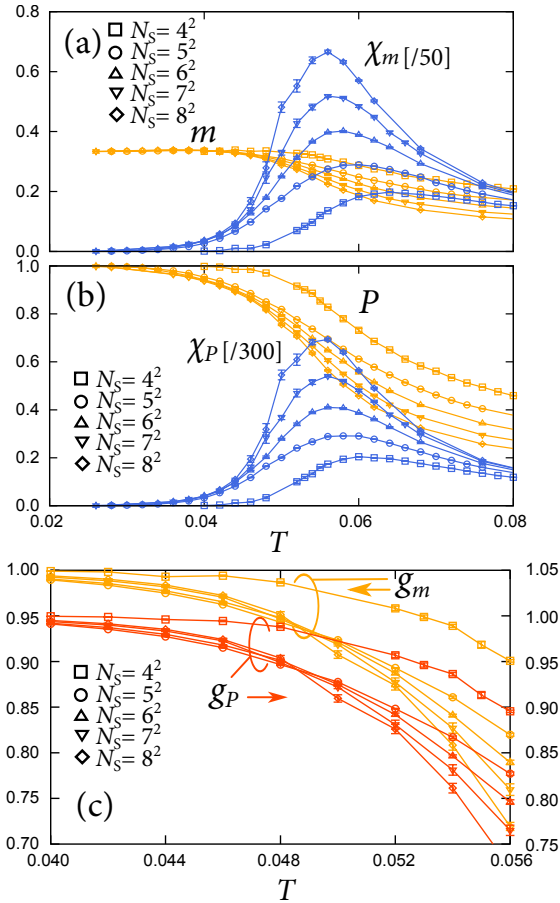}
   \caption{(color online).
   MC results for (a) $m$ and $\chi_m$, (b) $P$ and $\chi_P$, and (c) $g_m$ and $g_P$ for $N_{\rm s}=4^2$, $5^2$, $6^2$, $7^2$, and $8^2$ and at $n=0.83$.
   }
   \label{fig:mc_llJ6}
\end{figure}

For $0.8\lesssim n <8/9$, an exotic state appears in the intermediate-temperature range, which we call the LL state as shown in the phase diagram in Fig.~\ref{fig:ntdiag}. Figure~\ref{fig:mc_llJ6}(a) shows the temperature dependences of $m$ and $\chi_m$ for $n=0.83$ and $J=6$. As shown in the figure, $m$ rapidly increases at $T\sim 0.05$ with saturation to $1/3$. Accordingly, $\chi_m$ shows a divergent peak as increasing the system size. At the same time, as shown in Fig.~\ref{fig:mc_llJ6}(b), $P$ also increases rapidly to 1 and its susceptibility $\chi_P$ exhibits a divergent peak, indicating that most of the triangles become two-up one-down (or one-up two-down) coherently below $T\sim0.05$.

The Binder parameters for $m$ and $P$, $g_m$ and $g_P$, respectively, are shown in Fig.~\ref{fig:mc_llJ6}(c). Both show a crossing, indicating the transition is of second order. The critical temperatures determined from the two independent Binder analyses are in good accordance with each other; $T_c=0.051(4)$. However, this transition temperature is much higher than the transition to the $q=0$ FR state discussed in Sec.~\ref{ssec:sl}, which should be $T_c^{(q=0)}<0.028$. The results indicate that before entering the $q=0$ FR state the system exhibits another phase transition with a fractional magnetic moment $m\simeq 1/3$ at $T_c$.

\begin{figure}[tb]
   \includegraphics[width=\linewidth]{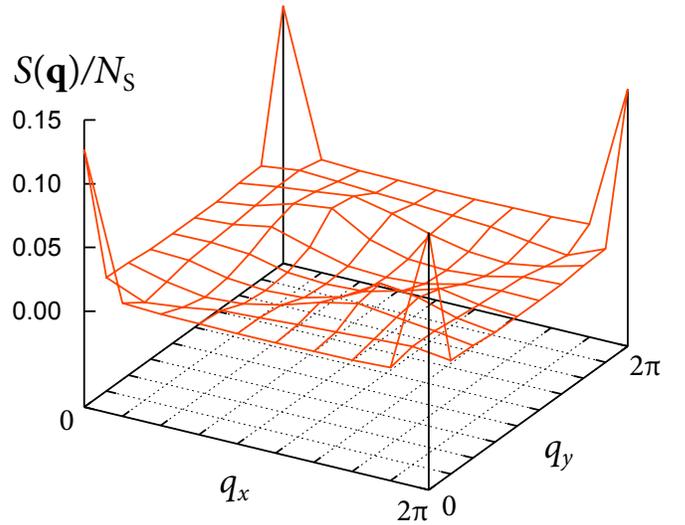}
   \caption{(color online).
   MC results of $S({\bf q})/N_{\rm s}$ at $n=0.84$ and $J=6$ for $T=0.03$.  The data are calculated for $N_{\rm s}=9^2$.
   }
   \label{fig:mc_sk84}
\end{figure}

Figure~\ref{fig:mc_sk84} shows the MC result of $S({\bf q})/N_{\rm s}$ at $T=0.03$ and $n=0.84$ calculated with $N_{\rm s}=9^2$ (Ref.~\onlinecite{note_figure}). The result shows a small Bragg peak at ${\bf q}={\bf 0}$, which corresponds to the fractional magnetic moment $m\simeq 1/3$. However, there are no other Bragg peaks indicating the formation of magnetic superstructure; it only shows two small humps and a node along $q_x=q_y$.

These results show that the intermediate state for $T_c^{(q=0)} < T < T_c$ is a peculiar state, in which all the triangles in the kagome lattice follow two-up one-down constraint, but still fluctuating between different spin configurations that satisfy the constraint. The spin state can be viewed in terms of the emergent degrees of freedom, self-avoiding up-spin loops, as schematically shown in Fig.~\ref{fig:kagome}(b). In this state, the up-spin loops are thermally fluctuating with exhibiting no LRO, and the down spins are isolated from each other by the fluctuating up-spin loops. Hence, we call this intermediate state the LL state.~\cite{Ishizuka2013c} In this loop picture, the two successive transitions at $T_c$ and $T_c^{(q=0)}$ correspond to the formation of loops and their crystallization, respectively.

\subsubsection{Electronic structure}\label{sssec:ll_ele}

\begin{figure}
   \includegraphics[width=\linewidth]{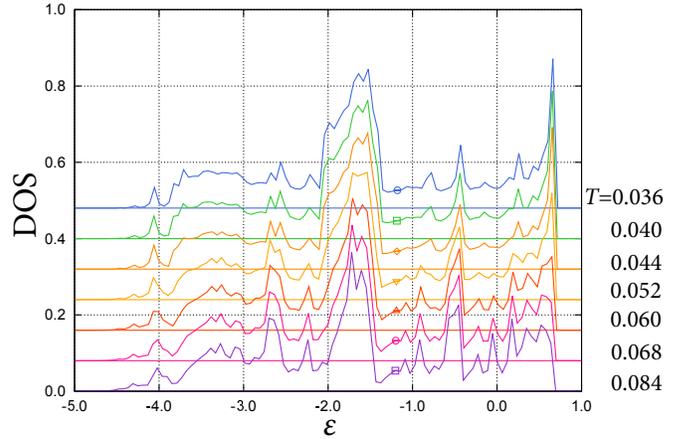}
   \caption{(color online).
   MC results for the DOS calculated while varying temperature at $n=0.83$ and $J=6$. The typical error bars are shown at $\varepsilon \sim -1.25$.
   }
   \label{fig:dos}
\end{figure}

Next, we look into the electronic structure of itinerant electrons in the LL state. Figure~\ref{fig:dos} shows the DOS at $n=0.83$ and $J=6$ with varying temperature calculated by the MC simulation for the current model. The Fermi level is set at $\varepsilon=0$. As shown in the figure, when the system enters into the LL state for $T < T_c = 0.051(4)$, a sharp peak develops at the upper edge of the band at $\varepsilon \sim 0.65$. This peak is ascribed to the confined electronic state associated with the up-spin loops in the LL state, as discussed below. 

In the LL state, a confinement of electrons in the up-spin loops takes place as the consequence of the quantum phase interference. This is explicitly seen by considering a wave function
\begin{eqnarray}
\left|\psi_{l_j(\{S_i\})}\right> = \sum_{i\in l_j(\{S_i\})} (-1)^{|i-j|}\left|i\right>.\label{eq:flatband}
\end{eqnarray}
Here, $l_j(\{S_i\})$ indicates the up-spin loop in the spin configuration $\{S_i\}$, $j$ is a site included in the loop $l_j$, $|i\rangle$ is the electronic state localized at $i$th site, and $|i-j|$ is the Manhattan distance between $i$th and $j$th sites. The sum is taken over all the sites in the loop $l_j$. The state in Eq.~(\ref{eq:flatband}) is an eigenstate of the Hamiltonian for the spin configuration $\{S_i\}$, as
\begin{eqnarray}
H(\{S_i\})\left|\psi_{l_j(\{S_i\})}\right>=(2-J-\mu)\left|\psi_{l_j(\{S_i\})}\right>.
\end{eqnarray}
For the results in Fig.~\ref{fig:dos}, the chemical potential is around $\mu\sim-4.65=-J+1.35$. Hence, the development of the sharp peak in the DOS at $\varepsilon \sim 0.65$ is ascribed to the loop formation in the LL state.

\subsubsection{Optical conductivity}\label{sssec:ll_opt}

\begin{figure}
   \includegraphics[width=\linewidth]{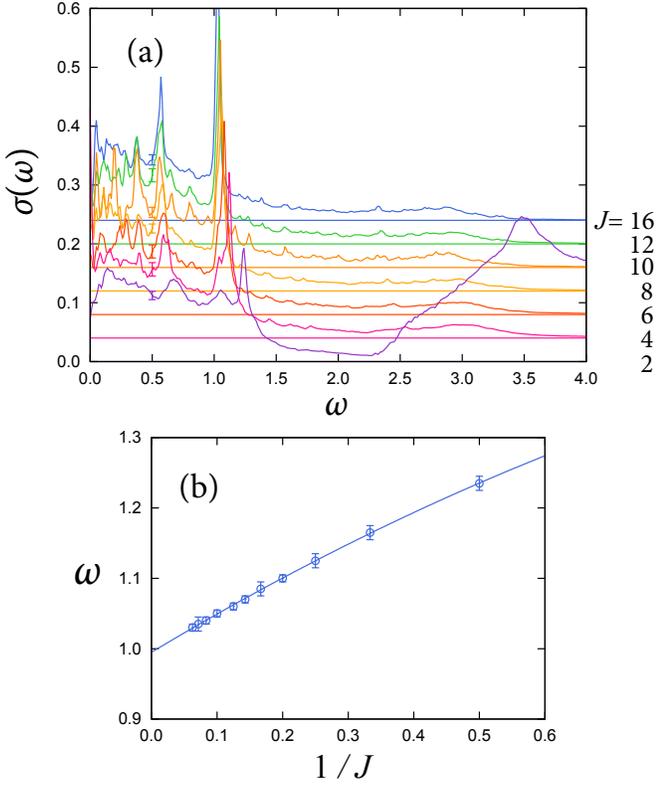}
   \caption{(color online).
   (a) Optical conductivity $\sigma(\omega)$ calculated by simple average over LL configurations while varying $J$ at $n=0.843$ for a $2^2$ supercell of $N=3\times 12^2$ sites. The typical error bars are shown at $\omega=0.5$. (b) $J$ dependence of the peak position of $\sigma(\omega)$ at $\omega \sim 1$. The dotted line shows the fitting by $\omega=0.995+0.558/J-0.155/J^2$.
   }
   \label{fig:sigma1}
\end{figure}

In addition to the peak structure of the DOS, another characteristic feature appears in the transport phenomena due to the quantum confinement. Figure~\ref{fig:sigma1} shows the result of optical conductivity $\sigma(\omega)$. Here, to extract the effect of characteristic spin correlations in the LL state, we calculate $\sigma(\omega)$ by taking simple average over different spin patterns in the ideal LL manifold, i.e., all the triangles satisfy the two-up one-down local constraint. The calculations were done by using the Kubo formula in Eq.~(\ref{eq:kubo_part}) for 24 different spin patterns. Figure~\ref{fig:sigma1}(a) is the result of $\sigma(\omega)$ calculated at $n = 0.843$ for various $J$. All the results show a sharp peak at $\omega=\omega_p\sim 1.0$-$1.2$, which shifts to lower $\omega$ for larger $J$.

The characteristic peak comes from the transition process between two confined states in the six-site loops. In the limit of $J\to\infty$, electrons are confined in the up-spin loops or at isolated down-spin sites as discussed above;~\cite{Jaubert2012} the contribution to $\sigma(\omega)$ comes only from the transition process between the electronic states in the same loop. Hence, sharp peaks appear in $\sigma(\omega)$ corresponding to the discrete energy levels in the finite-length loops. In the current kagome case, the most dominant loops are the shortest ones with the length of six sites. In the six-site loops, the energy difference between the unoccupied and occupied levels at this filling (the highest and second highest levels) is $t$. Hence, we expect a sharp peak at $\omega_p=1$ in the limit of $J \to \infty$. For large but finite $J$, as the highest energy state is the state described by the wave function in Eq.~(\ref{eq:flatband}), this state remains the same as that of $J=\infty$. On the other hand, for the second highest level, the hybridization to the localized state at down-spin sites shifts the eigenenergy to a lower energy. Hence, it is expected that the peak in $\sigma(\omega)$ shifts to a higher $\omega$ as decreasing $J$. This is confirmed by the fitting shown in Fig.~\ref{fig:sigma1}(b).

Interestingly, the peak persists in the weak $J$ region where the exchange splitting $2J$ is comparable or even smaller than the bare bandwidth $6t$ and the above strong-$J$ argument appears to be no longer valid. In a recent study on a metal-insulator transition caused by correlated potentials, a LL-type local correlation induces a metal-insulator transition at a considerably smaller potential than the bandwidth by confining the electrons in the loops.~\cite{Ishizuka2011} The persisting resonant peak in $\sigma(\omega)$ is likely to be the consequence of this confinement.

\begin{figure}
   \includegraphics[width=\linewidth]{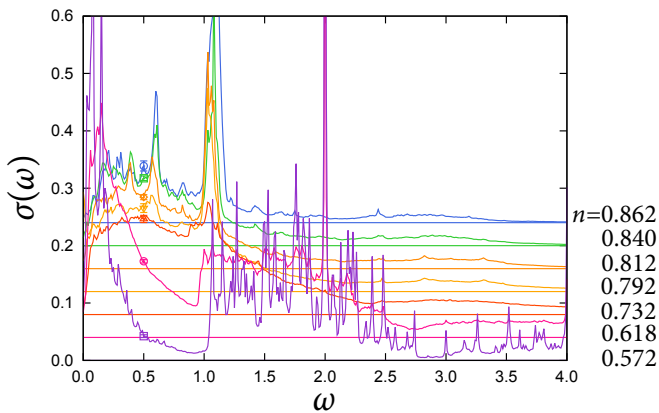}
   \caption{(color online).
   Optical conductivity $\sigma(\omega)$ calculated by MC simulation while varying $n$ at $J=6$ for a $4^2$ supercell of $N=3\times 6^2$ sites at $T=0.04$.
   The scattering rate in the Kubo formula is taken as $\tau^{-1}=0.01$.
   The typical error bars are shown at $\omega=0.5$.
   }
   \label{fig:sigma2}
\end{figure}

Emergence of the characteristic peak is also observed in the thermodynamic average obtained by the MC simulation. Figure~\ref{fig:sigma2} shows the MC result of $\sigma(\omega)$ while varying $n$ at $T=0.04$ and $J=6$. With increasing $n$ from the FM region, the peak at $\omega\sim 1$ shows sharp development for $n \gtrsim 0.8$. This shows that the resonant peak in the optical conductivity can be used as a sensitive signal to detect the LL state.

\section{Discussions and Summary}\label{sec:summary}

To summarize, we mapped out the phase diagram of the Ising-spin Kondo lattice model on a kagome lattice, using a Monte Carlo simulation at finite temperature, as well as variational calculations for the ground state. We presented that this model shows the rich phase diagram with various magnetic states induced by thermal fluctuations: the partially disordered, Kosterlitz-Thouless-like, partially ferromagnetic, and loop-liquid states. We also discussed phase transitions and crossovers to the competing phases including the conventional magnetically ordered states, such as the ferromagnetic and ferrimagnetic states.

In addition, we studied the electronic structure and optical conductivity in the partially disordered and loop-liquid phases. In the partially disordered state, we showed that the electronic density of states exhibits characteristic spikes, which originate from the confined electronic states on the strings and loops of the same spin sites. This is related to the geometry of kagome lattice, and hence, the spikes are a characteristic feature of the kagome model that is absent in the triangular lattice case. We also showed that a charge gap opens in the electronic density of states in the partial disorder; this phase is an insulator. For the loop-liquid state, we found a sharp peak at the upper edge of the energy band in the density of states, which originates from the presence of closed loops in the thermally fluctuating spin configurations. Also, we showed that a related resonant peak appears in the optical conductivity. These results imply that the electronic states of itinerant electrons are strongly affected by the underlying spin textures in the loop-liquid state. We also note that the arguments presented in this paper are based on a simple real-space picture. Hence, though the electronic structure of the actual materials are often much more complicated than the model considered here, we believe similar behavior can be present in the materials with multi-orbital itinerant electrons. Hence, our results might provide useful experimental probes for the exotic thermally-induced phases.

Among the phases we found, the partially disordered and loop liquid states can be viewed as a partial lifting of the degenerated ground state manifold, where the spins fluctuate only in a subspace of the manifold of kagome Ising antiferromagnet. In the frustrated magnets, subdominant interactions, e.g., further-neighbor interactions, tend to lift the ground state degeneracy, and often believed to drive magnetic long-range orders. In contrast, the results presented here are counter examples in which the ``cooperation" between conflicting two factors, thermal fluctuations and degeneracy-lifting interactions, can give birth to rich phenomena by partially lifting the degenerated ground states.

\acknowledgements

The authors thank N. Furukawa, L. D. C. Jaubert, T. Misawa, and K. Penc for fruitful discussions. This research was supported by KAKENHI (No. 24340076), the Strategic Programs for Innovative Research (SPIRE), MEXT, and the Computational Materials Science Initiative (CMSI), Japan. HI is supported by Grant-in-Aid for JSPS Fellows and JSPS Postdoctoral Fellowships for Research Abroad.

\appendix
\section{Results for Ferromagnetic Phases}\label{ssec:fm&pfm} 

\begin{figure}
   \includegraphics[width=.85\linewidth]{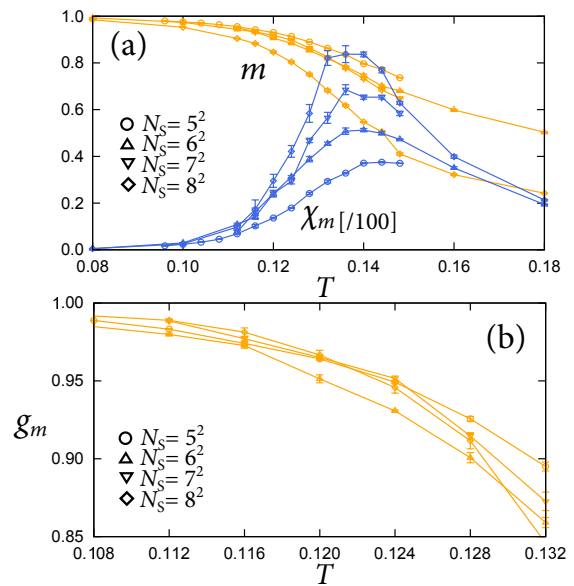}
   \caption{(color online).
   MC results for (a) $m$ and $\chi_m$, and (b) $g_m$ at $n=0.52$ and $J=6$. 
   The data are calculated for $N_{\rm s}=5^2$, $6^2$, $7^2$, and $8^2$.
   }
   \label{fig:mc_fmJ6}
\end{figure}

In this appendix, we present the MC results of the ferromagnetic phases in Figs.~\ref{fig:diagram_n23} and \ref{fig:ntdiag}. We start from small $n$ region of Fig.~\ref{fig:ntdiag}. In this region, the FM phase appears dominantly in the phase diagram, with $m$ approaching its saturated value $1$ in the low-temperature limit. This phase is connected to the FM state in the  small $n$ and weak $J$ limit stabilized by effective ferromagnetic RKKY interactions, and also to that in the large $J$ limit stabilized by the ferromagnetic double-exchange interactions.~\cite{Zener1951,Anderson1955}

The typical temperature dependence of $m$ and $\chi_m$ is shown at $n=0.52$ and $J=6$ in Fig.~\ref{fig:mc_fmJ6}(a). The result shows a rapid increase in $m$ with decreasing temperature. Correspondingly, $\chi_m$ exhibits a peak whose height increases as the system size increases. These indicate a magnetic transition to the FM phase. The transition temperature is determined by the Binder parameter~\cite{Binder1981} shown in Fig.~\ref{fig:mc_fmJ6}(b).

\begin{figure}
   \includegraphics[width=.85\linewidth]{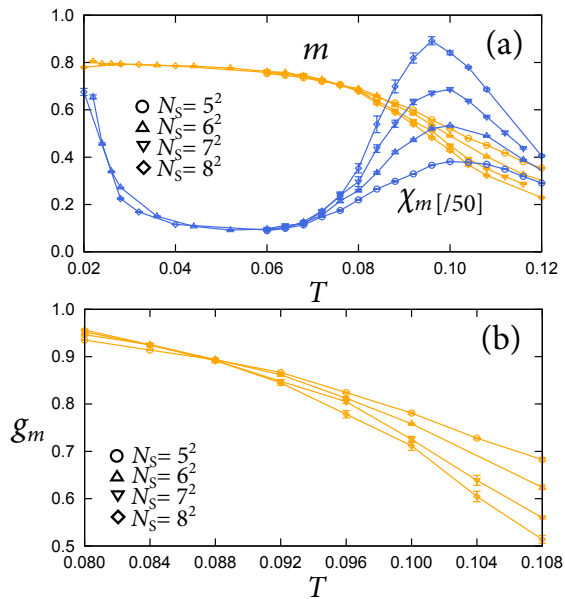}
   \caption{(color online).
   MC results for (a) $m$ and $\chi_m$, and (b) $g_m$ at $n=0.64$ and $J=6$. The data are calculated for $N_{\rm s}=5^2$, $6^2$, $7^2$, and $8^2$.
   }
   \label{fig:mc_pfmJ6}
\end{figure}

As the electron density increases, the saturation value of $m$ decreases. Figure~\ref{fig:mc_pfmJ6}(a) shows MC results for $m$ and $\chi_m$ at $n=0.64$. The overall behavior of $m$ looks similar to that for $n=0.52$ in Fig.~\ref{fig:mc_fmJ6}(a). $\chi_m$ also shows divergent behavior at the onset of the rapid increase in $m$. $T_c$ is estimated to be 0.088(6) by the Binder parameter shown in Fig.~\ref{fig:mc_fmJ6}(b). However, for $T\ll T_c$, $m$ appears to saturate at $\sim0.8$, not $1.0$. The magnetization smaller than $1$ suggests the possibility of magnetic superlattice structure, such as FR orders. However, $\chi_m$ increases again at low temperature well below $T_c$. This behavior is not expected for FR orders.

\begin{figure}[tb]
   \includegraphics[width=.9\linewidth]{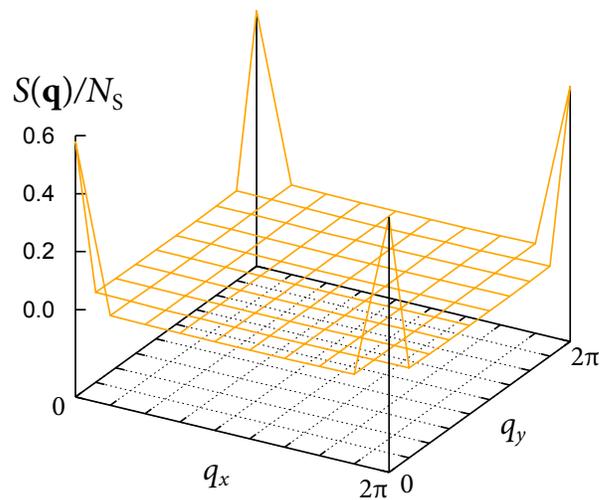}
   \caption{(color online).
   MC results of $S({\bf q})/N_{\rm s}$ at $n=0.65$ and $J=6$ for $T=0.03$. The data are calculated for $N_{\rm s}=9^2$.
   }
   \label{fig:mc_sk65}
\end{figure}

The absence of such long-period magnetic structure is also confirmed by the calculation of the spin structure factor $S({\bf q})$. Figure~\ref{fig:mc_sk65} shows the result of $S({\bf q})/N_{\rm s}$ at $T=0.03$ and $N_{\rm s}=9^2$ (Ref.~\onlinecite{note_figure}). The result shows a Blagg peak only at ${\bf q}=\bm{0}$, that corresponds to the net magnetic moment. There are no other peaks in $S({\bf q})$, indicating that the low-$T$ phase is the FM state with a smaller saturation moment than $m=1$. We call this state the PFM state.

For the current model with Ising localized spins, the PFM state is expected to become unstable in the low-temperature limit, as the reduction of $m$ should be induced by thermal fluctuations. In the MC simulation, however, we could not find any indication of the instability at a lower temperature; our MC results merely show the freezing of sampling in the low temperature region. On the other hand, a variational calculation comparing the ground state energy for different magnetic ordered states suggests that the ground state for this parameter is the phase separation between the FM and $q=0$ FR states, as indicated in the bottom strip of Fig.~\ref{fig:ntdiag}. From these facts, in the low temperature region where our MC simulation cannot reach, the PFM state is most likely to be taken over by the phase separation between the FM and $q=0$ FR states.

\section{Results for Ferrimagnetic Phases} 
\label{ssec:sl}

\begin{figure}
   \includegraphics[width=\linewidth]{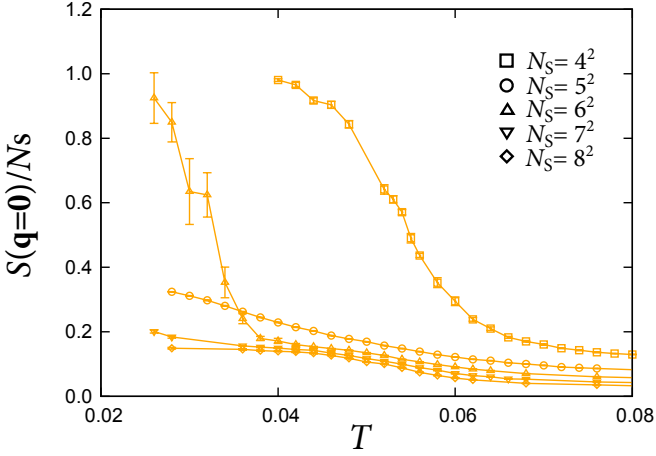}
   \caption{(color online).
   The temperature dependence of $S({\bf q}={\bf 0})/N_{\rm s}$ at $n=0.83$ and $J=6$. The data are calculated for and $N_{\rm s}=4^2$, $5^2$, $6^2$, $7^2$, and $8^2$.
   }
   \label{fig:mc_llJ6_sk}
\end{figure}

In this appendix, we present the MC results of the ferrimagnetic phases in \ref{fig:ntdiag}. In the high density region ($n\gtrsim0.8$) of Fig.~\ref{fig:ntdiag}, a phase transition to a FR ordered state is found both in the MC and variational calculations (see Fig.~\ref{fig:ntdiag}). Figure~\ref{fig:mc_llJ6_sk} shows the MC result of $S({\bf q}={\bf 0})/N_{\rm s}$ for $n=0.83$ at $J=6$. The result shows a rapid increase in $S({\bf q}={\bf 0})/N_{\rm s}$ to 1 for $N_{\rm s}=4^2$ and $6^2$. On the other hand, the net magnetization approaches $m=1/3$ at low temperature, as shown in Fig.~\ref{fig:mc_llJ6}. These results imply an instability toward the $q=0$ FR ordered state shown in Fig.~\ref{fig:kagome}(d).

As shown in Fig.~\ref{fig:mc_llJ6_sk}, the onset temperature decreases for larger $N_{\rm s}$ although the results show strong finite-size effects with different behavior for even and odd $N_{\rm s}$. We could not observe the upturn of $S({\bf q}={\bf 0})$ for  larger sizes $N_{\rm s} \ge 7^2$. Hence, numerically, we could not determine the critical temperature for the transition toward the $q=0$ FR ordered state. However, as the instability is consistently seen in the result of variational calculation as shown in the bottom of Fig.~\ref{fig:ntdiag}, the ground state at $n \simeq 0.83$ is likely to be the $q=0$ FR state with the critical temperature $T_c^{(q=0)}\le0.028$.

\begin{figure}
   \includegraphics[width=0.8\linewidth]{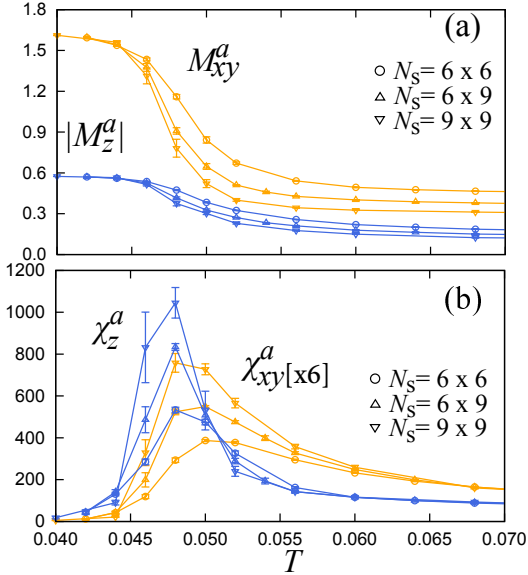}
   \caption{(color online).
   MC results for (a) $M_{xy}^a$ and $|M_z^a|$, and (b) $\chi_{xy}^a$ and $\chi_{z}^a$ at $n=8/9$ and $J=6$. The data are calculated for $N_{\rm s}= 6 \times 6$, $6\times 9$, and $ 9\times 9$.
   }
   \label{fig:mc_n89}
\end{figure}

On the other hand, at $n=8/9$, the system shows phase transition to a different FR ordered state. Figure~\ref{fig:mc_n89} shows MC results for $J=6$ and $n=8/9$. As shown in Fig.~\ref{fig:mc_n89}(a), both $M_{xy}^a$ and $|M_z^a|$ show a rapid increase at $T\sim 0.05$. Correspondingly, the susceptibilities show a divergent increase with increasing $N_{\rm s}$. The transition temperature is estimated to be around $T_c=0.046(2)$ from the extrapolation of the peaks in $\chi^a_{xy}$. The behavior indicates the phase transition to the $\sqrt3\times\sqrt3$ FR ordered state [see Fig.~\ref{fig:kagome}(c)]. These results are consistent with the variational phase diagram shown in the bottom of Fig.~\ref{fig:ntdiag}.

\begin{figure}[tb]
   \includegraphics[width=\linewidth]{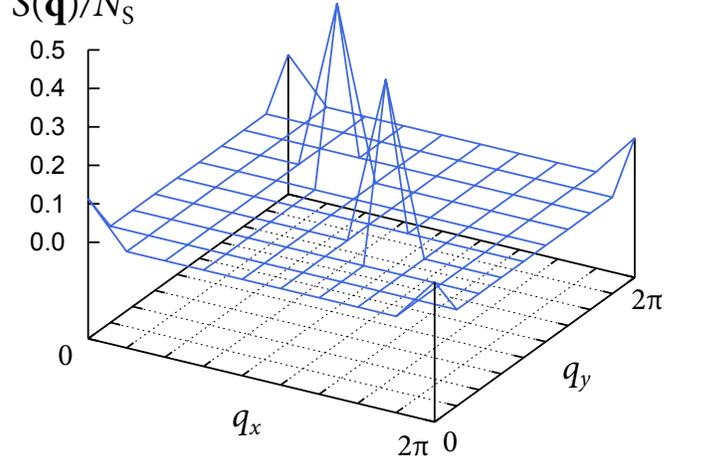}
   \caption{(color online).
   MC results of $S({\bf q})/N_{\rm s}$ at $n=8/9$ and $J=6$ for $T=0.03$. The data are calculated for $N_{\rm s}=9^2$.
   }
   \label{fig:mc_sk89}
\end{figure}

The $\sqrt3\times\sqrt3$ FR order is more clearly seen in the spin structure factor. Figure~\ref{fig:mc_sk89} shows the result of $S({\bf q})/N_{\rm s}$ at $T=0.03$.~\cite{note_figure} The result shows three Bragg peaks: ${\bf q}={\bf 0}$ corresponding to the net magnetization, and ${\bf q}=(4\pi/3,2\pi/3)$ and $(2\pi/3,4\pi/3)$ to the three-sublattice magnetic superstructure.



\begin{references}
\bibitem[$\ast$]{KITP} Present Address: Kavli Institute for Theoretical Physics, University of California, Santa Barbara, California 93106, USA.
\setcounter{enumiv}{0}
\bibitem{Diep2004}       For a recent review, see: {\it Frustrated Spin Systems}, edited by H. T. Diep (World Scientific Publishing, 2004).
\bibitem{Lacroix2011}    For a recent review, see: {\it Introduction to Frustrated Magnetism}, edited by C. Lacroix, P. Mendels, and F. Mila (Springer-Verlag, 2011).
\bibitem{Pauling1935}    L. Pauling, J. Am. Chem. Soc. {\bf 57}, 2680 (1935).
\bibitem{Wannier1950}    G. H. Wannier, Phys. Rev. {\bf 79}, 358 (1950).
\bibitem{Houtappel1950}  R. M. F. Houtappel, Physica {\bf 16}, 425 (1950).
\bibitem{Takayama1983}     H. Takayama, K. Matsumoto, H. Kawahara, and K. Wada, J. Phys. Soc. Jpn. {\bf 52}, 2888 (1983).
\bibitem{Fujiki1984}       S. Fujiki, K. Shutoh, and S. Katsura, J. Phys. Soc. Jpn. {\bf 53}, 1371 (1984).
\bibitem{Fujiki1986}       S. Fujiki, K. Shutoh, S. Inawashiro, Y. Abe, and S. Katsura, J. Phys. Soc. Jpn. {\bf 55}, 3326 (1986).
\bibitem{Takagi1993}       T. Takagi and M. Mekata, J. Phys. Soc. Jpn. {\bf 62}, 3943 (1993).
\bibitem{Takagi1995}       T. Takagi and M. Mekata, J. Phys. Soc. Jpn. {\bf 64}, 4609 (1995).
\bibitem{Mekata1977}       M. Mekata, J. Phys. Soc. Jpn. {\bf 42}, 76 (1977).
\bibitem{Todoroki2004}     N. Todoroki and S. Miyashita, J. Phys. Soc. Jpn. {\bf 73}, 412 (2004).
\bibitem{Yoshida2011}      H. Yoshida, E. Takayama-Muromachi, and M. Isobe, J. Phys. Soc. Jpn. {\bf 80}, 123703 (2011).
\bibitem{Matsuda2012}      M. Matsuda, C. de la Cruz, H. Yoshida, M. Isobe, and R. S. Fishman, Phys. Rev. B {\bf 85}, 144407 (2012).
\bibitem{Takatsu2010}      H. Takatsu, S. Yonezawa, S. Fujimoto, and Y. Maeno, Phys. Rev. Lett {\bf 105}, 137201 (2010).
\bibitem{Takatsu2014}      H. Takatsu, G. Nenert, H. Kadowaki, H. Yoshizawa, M. Enderle, S. Yonezawa, Y. Maeno, J. Kim, N. Tsuji, M. Takata, Y. Zhao, M. Green, C. Broholm, Phys. Rev. B {\bf 89}, 104408 (2014).
\bibitem{Ruderman1954}     M. A. Ruderman and C. Kittel, Phys. Rev. {\bf 96}, 99 (1954).
\bibitem{Kasuya1956}       T. Kasuya, Prog. Theor. Phys. {\bf 16}, 45 (1956).
\bibitem{Yosida1957}       K. Yosida, Phys. Rev. {\bf 106}, 893 (1957).
\bibitem{Ishizuka2012}     H. Ishizuka and Y. Motome, Phys. Rev. Lett. {\bf 108}, 257205 (2012).
\bibitem{Ishizuka2013}     H. Ishizuka and Y. Motome, Phys. Rev. B {\bf 87}, 155156 (2013).
\bibitem{Martin2008}       I. Martin and C. D. Batista, Phys. Rev. Lett. {\bf 101}, 156402 (2008).
\bibitem{Akagi2010}        Y. Akagi and Y. Motome, J. Phys. Soc. Jpn. {\bf 79}, 083711 (2010).
\bibitem{Kumar2010}        S. Kumar and J. van den Brink, Phys. Rev. Lett. {\bf 105}, 216405 (2010).
\bibitem{Kato2010}         Y. Kato, I. Martin, and C. D. Batista, Phys. Rev. Lett. {\bf 105}, 266405 (2010).
\bibitem{Dunlap1990}       D. H. Dunlap, H.-L. Wu, and P. W. Phillips, Phys. Rev. Lett. {\bf 65}, 88 (1990).
\bibitem{Ishizuka2011}     H. Ishizuka, M. Udagawa, and Y. Motome, Phys. Rev. B {\bf 83}, 125101 (2011).
\bibitem{Ishizuka2013d}    H. Ishizuka and Y. Motome, Phys. Rev. B {\bf 87}, 081105 (2013).
\bibitem{Chern2012}        G.-W. Chern, A. Rahmani, I. Martin, and C. D. Batista, arXiv:1212.3617.
\bibitem{Ishizuka2013b}    H. Ishizuka and Y. Motome, J. Korean Phys. Soc. {\bf 63}, 579 (2013).
\bibitem{Ishizuka2013c}    H. Ishizuka and Y. Motome, Phys. Rev. B {\bf 88}, 081105 (2013).
\bibitem{Yunoki1998}       S. Yunoki, J. Hu, A. L. Malvezzi, A. Moreo, N. Furukawa, and E. Dagotto, Phys. Rev. Lett. {\bf 80}, 845 (1998).
\bibitem{Rahman1972}       A. Rahman and F. H. Stillinger, J. Chem. Phys. {\bf 57}, 4009 (1972).
\bibitem{Barkema1998}      G. T. Barkema and M. E. J. Newman, Phys. Rev. E {\bf 57}, 1155 (1998).
\bibitem{Melko2001}        R. G. Melko, B. C. den Hertog, and M. J. P. Gingras, Phys. Rev. Lett. {\bf 87}, 067203 (2001).
\bibitem{Ozeki2003}        Y. Ozeki, K. Kasono, N. Ito, and S. Miyashita, Physica A {\bf 321}, 271 (2003).
\bibitem{Jaubert2012}      L. D. C. Jaubert, S. Piatecki, M. Haque, and R. Moessner, Phys. Rev. B {\bf 85}, 054425 (2012).
\bibitem{note_figure}      The plot range of the spin structure factor was incorrect in Figs.~2(c)-2(e) in Ref.~\onlinecite{Ishizuka2013c}: the plots were for $0 \le k_x \le 4\pi/3$ and $0 \le k_y \le 4\pi/3$.
\bibitem{Binder1981}       K. Binder, Z. Phys. B {\bf 43}, 119 (1981).
\bibitem{Zener1951}        C. Zener, Phys. Rev. {\bf 82}, 403 (1951).
\bibitem{Anderson1955}     P. W. Anderson and H. Hasegawa, Phys. Rev. {\bf 100}, 675 (1955).
\end{references}
\end{document}